\documentclass[a4paper,11pt]{article}
\usepackage{pos}
\usepackage{braket,here}
\title{Theoretical aspects of $D^0-\bar{D}^0$ mixing}

\author*{Hiroyuki Umeeda}

\affiliation{Institute of Physics, Academia Sinica,\\
Taipei 11529, Taiwan, Republic of China}

\emailAdd{umeeda@gate.sinica.edu.tw}

\abstract{
Observables in the $D^0-\bar{D}^0$ mixing can be theoretically analyzed by the operator product expansion (OPE), in which $1/m_c$ is regarded as an expansion parameter. Since the contributions of four-quark operators are strongly suppressed by the Glashow-Iliopoulos-Maiani (GIM) mechanism, the order of magnitude of the width difference is still not reproduced in the OPE analysis. In view of this issue, quark-hadron duality, an assumption that is tacitly made in the OPE, is studied for the $D^0-\bar{D}^0$ mixing. In particular, the exclusive width difference and the inclusive counterpart can be compared within the 't Hooft model, two-dimensional QCD in the large-$N_c$ limit. It is shown that the order of magnitude of the exclusive width difference is enhanced relative to the 4D-like inclusive contributions of the four-quark operators, that is qualitatively consistent with the realistic observation of the $D^0-\bar{D}^0$ mixing.
}

\FullConference{%
  11th International Workshop on the CKM Unitarity Triangle (CKM2021)\\
  22-26 November 2021\\
  The University of Melbourne, Australia
}

\begin{document}
\maketitle
\section{Introduction}
Even though the theory of heavy quark has been well-established, charm physics is still regarded as a challenging topic. The two-fold difficulty in treating charm originates from (1) not sufficiently large mass of charm quark and (2) the strong cancellation in the presence of the Glashow-Iliopoulos-Maiani (GIM) mechanism \cite{Glashow:1970gm}. The $D^0-\bar{D}^0$ mixing, which occurs through $\Delta C=2$ interaction, is one of the processes that are made complicated by (1) and (2). In the literature, theoretical methods to analyze this process are broadly classified into two categories: the exclusive and inclusive approaches. For the former cases, where recent analyses were performed in Refs.~\cite{Cheng:2010rv, Gronau:2012kq, Jiang:2017zwr}, the order of the magnitude of $y=\Delta \Gamma_{D^0}/2\Gamma_{D^0}$ is properly reproduced, while there exists difficulty in including multi-body channels. Meanwhile, for the inclusive case, which relies on the operator product expansion (OPE), the status is rather different: the theoretical predictions based on the next-to-leading order precision and an average of experimental data give,
\begin{eqnarray}
y_{\rm th} &=& 6\times 10^{-7},\quad\qquad\quad\quad\quad\;\;\;\:\:\:  (\mathrm{OPE}~[5])\\
y_{\rm th} &\leq& 4.7\times 10^{-7}\ldots 1.6\times 10^{-6},\quad (\mathrm{OPE}~[6])\\
y_{\rm ex}&=& (0.63\pm 0.07)\%.\quad\qquad\quad\:\:\:\: (\mathrm{HFLAV}~\mathrm{no}~\mathrm{CPV}~[7])
\end{eqnarray}
Thus, the OPE results that stem from the contributions of the four-quark operators do not reproduce the order of magnitude of the experimental data. The suppression of the theoretical results is caused by the aforementioned GIM mechanism, that strongly works presumably only for the four-quark operators \cite{Georgi:1992as,Ohl:1992sr,Bigi:2000wn,Falk:2001hx,Bobrowski:2010xg}. See Refs.~\cite{Li:2020xrz, Lenz:2020efu} for the recent works to tackle this issue.
\par
Regarding the OPE analysis, it should be noted that quark-hadron duality is tacitly assumed so that violation of duality potentially causes theoretical uncertainty. In order to investigate this aspect, the 't Hooft model \cite{tHooft:1974pnl} offers a qualitative testing ground of QCD. Since this is a solvable model, we can determine the exclusive width difference unambiguously within the formalism. The obtained result is compared with the OPE prediction from the four-quark operators to check how reliable the inclusive theoretical estimation for the $D^0-\bar{D}^0$ mixing is \cite{Umeeda:2021llf} (see also Ref.~\cite{Umeeda:2021gfn} for the recent work for heavy meson lifetimes).
\section{Formalism in the 't Hooft model}
For a meson that consists of $q_1\bar{q}_2$, the Bethe-Salpeter equation in the light-cone gauge, referred to as the 't Hooft equation, is given by,
\begin{eqnarray}
M_k^2\phi_k^{(q_1\bar{q}_2)}(x)=\left(\frac{m_1^2-\beta^2}{x}+\frac{m_2^2-\beta^2}{1-x}\right)\phi_k^{(q_1\bar{q}_2)}(x)-\beta^2\mathrm{Pr}\int_{0}^1\mathrm{d}y\frac{\phi_k^{(q_1\bar{q}_2)}(x)}{(x-y)^2}\label{Eq:1}
\end{eqnarray}
where $m_{1}$ and $m_{2}$ are bare masses of $q_1$ and $\bar{q}_2$, respectively. The notation of the QCD coupling is introduced by $\beta^2=g^2 N_c/(2\pi)$, which is fixed so as to fit the string tension of QCD$_4$, leading to $\beta= 340~$MeV. In Eq.~(\ref{Eq:1}), $k=0$ represents the equation for a ground state while ones for $k=1, 2,\cdots$ are associated with radially excited states.  $M_k~ (\phi_k^{(q_1\bar{q}_2)})$ is a meson mass (light-cone wave function) of the $k-$th radial state. In this work, we numerically solve the 't Hooft equation by means of the BSW-improved Multhopp technique \cite{Brower:1978wm}, where the light-cone wave function is expanded as a series of the trigonometric function. The detail of this method is summarized, {\it e.g.}, in Ref.~\cite{Lebed:2000gm}.\par
The matrix element of the axial vector current for a $k-$th radial state is defined by,
\begin{eqnarray}
\bra{0}\bar{q}_2\gamma_\mu\gamma_5 q_1\ket{H_k(p)}=\sqrt{\frac{N_c}{\pi}}c_k^{(q_1\bar{q_2})}p_\mu,\quad c_k^{(q_1\bar{q_2})}=\int_0^1 \mathrm{d}x \phi_k^{(q_1\bar{q}_2)}(x),
\end{eqnarray}
where $c_k^{(q_1\bar{q_2})}$ denotes the normalized decay constant of the relevant meson.
\section{$D^0-\bar{D}^0$ mixing}
\label{Sec:3}
In the CP conserving limit, we define an absorptive part of the matrix element of the $D^0\to \bar{D}^0$ transition amplitude for both exclusive and inclusive cases as follows,
\begin{eqnarray}
\Delta \Gamma_{D^0}^{(\alpha)}=\frac{\bra{\bar{D}^0}\mathcal{H}_{\rm abs}^{(D^0)}\ket{D^0}}{M_{D^0}},\quad \alpha=\mathrm{exc},\: \mathrm{inc}\label{Eq:6}.
\end{eqnarray}
The contributions to Eq.~(\ref{Eq:6}) consist of three parts depending on intermediate flavors,
\begin{eqnarray}
\Delta \Gamma_{D^0}^{(\alpha)}/2&=&\lambda_d^2\Gamma_{dd}^{(D^0,\: \alpha)}+2\lambda_s \lambda_d\Gamma_{sd}^{(D^0,\: \alpha)}+\lambda_s^2\Gamma_{ss}^{(D^0,\: \alpha)}\nonumber\\
&=&\lambda_s^2\left[\Gamma_{dd}^{(D^0,\: \alpha)}+\Gamma_{ss}^{(D^0,\: \alpha)}-2\Gamma_{sd}^{(D^0,\: \alpha)}\right]+2\lambda_s \lambda_b\left[\Gamma_{dd}^{(D^0,\: \alpha)}-\Gamma_{sd}^{(D^0,\: \alpha)}\right]+\lambda_b^2\Gamma_{dd}^{(D^0,\: \alpha)},\label{Eq:7}
\end{eqnarray}
with $\lambda_i $ being $V_{ci}^*V_{ui}~(i=d, s, b)$. In the second line of Eq.~(\ref{Eq:7}), the unitary relation given by $\lambda_d+\lambda_s+\lambda_b=0$ is used. It is worth noting that the width difference is characterized by SU(3) breaking since the first term proportional to $\lambda_s^2$ gives a sizable contribution due to $|\lambda_s|\gg |\lambda_b|$.
\par
In Secs.~\ref{Sec:3.1} and \ref{Sec:3.2}, detailed expressions of $\Gamma_{ij}^{(D^0,\:\alpha)}$ for $\alpha=\mathrm{exc}, \mathrm{inc}$ are obtained, respectively. For both analyses, we adopt a common weak interaction vertex with the generalized Lorentz structure that has a form of $(-ig_2/\sqrt{2})V_{\rm CKM}\gamma^\mu(c_{\rm V}+c_{\rm A}\gamma_5)$.
\subsection{Exclusive width difference}\label{Sec:3.1}
The exclusive result is obtained by summing over all the kinematically allowed  $D^0\to H_k H_m\to\bar{D}^0$ transitions, where $k$ and $m$ stand for radial excitation numbers. With $p_{km}$ being a momentum carried by either daughter meson, the exclusive sum of the width difference is given for $(i, j)=(d, d), (s, d), (s, s)$,
\begin{eqnarray}
\Gamma_{ij}^{(D^0,\: \mathrm{exc})}&=&\frac{4N_cG_F^2}{\pi}(c_{\rm V}^2-c_{\rm A}^2)^2\displaystyle\sum_{k, m}(-1)^{k+m}\frac{T^{(k, m)}_{(c\bar{u})(i, j)}T^{(m, k)\: *}_{(c\bar{u})(j, i)}}{2M_{D^0}|p_{km}|},\label{Eq:12}\\
T_{(Q\bar{q})(i, j)}^{(k, m)}&=&c_k^{(q\bar{i})}[(-1)^{k+1}M_k^2\mathcal{C}_m+m_Qm_j \mathcal{D}_m].\label{Eq:13}
\end{eqnarray}
In Eq.~(\ref{Eq:13}), $T_{(Q\bar{q})(i, j)}^{(k, m)}$ is the color-allowed tree diagram, for which the expression is obtained in Ref.~\cite{Grinstein:1997xk}. $\mathcal{C}_m$ and $\mathcal{D}_m$ represent overlap integrals explicitly given in Refs.~\cite{Bigi:1999fi, Umeeda:2021llf} up to the normalization. In order to guarantee the numerical stability of the exclusive width, in the above expression, the contribution of the triple overlap integral calculated in Ref.~\cite{Grinstein:1997xk} is neglected. This contribution is suppressed by at least $1/m_c^2$ \cite{Bigi:1999fi} so that its numerical impact is less pronounced as charm quark gets heavier.
\subsection{Inclusive width difference}\label{Sec:3.2}
In this case, the contributions of individual flavors on r.h.s. of Eq.~(\ref{Eq:7}) are obtained by calculating the box diagram in two-dimensions. For $(i, j)=(d, d), (s, d), (s, s)$, the result \cite{Umeeda:2021llf} reads,
\begin{eqnarray}
\Gamma_{ij}^{(D^0,\: \mathrm{inc})}=\frac{4N_cG_F^2}{\pi}(c_{\rm V}^2-c_{\rm A}^2)^2\left\{\left[F_{ij}^{\rm (th)}+2G_{ij}^{\rm (th)}\right]-\left[G_{ij}^{\rm (th)}+2H_{ij}^{\rm (th)}\right]R\right\}\left[c^{(c\bar{u})}_0\right]^2M_{D^0},\label{Eq:8}
\end{eqnarray}
where $R$ is represented by $[M_{D^0}/(m_c+m_u)]^2$ while $G_F$ represents a dimensionless Fermi constant. It is evident that $\Gamma_{ij}^{(D^0,\: \mathrm{inc})}$ is proportional to $N_c$ as well as the exclusive counterpart in Eq.~(\ref{Eq:12}) in this large-$N_c$ analysis. In order to compare the exclusive and inclusive width differences consistently, only the terms proportional to $(c_{\rm V}^2-c_{\rm A}^2)^2$ are considered in Eq.~(\ref{Eq:8}) while other terms proportional to $(c_{\rm V}^4-c_{\rm A}^4)$ are generically possible \cite{Umeeda:2021llf}. In Eq.~(\ref{Eq:8}), $F_{ij}^{\rm (th)}, G_{ij}^{\rm (th)}, H_{ij}^{\rm (th)}$ are phase space functions defined by ($z_\alpha=m_\alpha^2/m_c^2$ for $\alpha=i, j$),
\begin{eqnarray}
F_{ij}^{\rm (th)}=\sqrt{1-2(z_i+z_j)+(z_i-z_j)^2},\;\;
G_{ij}^{\rm (th)}=[z_i+z_j-(z_i-z_j)^2]/F_{ij}^{\rm (th)},\;\;
H_{ij}^{\rm (th)}=\sqrt{z_i z_j}/F_{ij}^{\rm (th)}.
\end{eqnarray}
Although Eq.~(\ref{Eq:8}) is obtained in two-dimensional spacetime, $F_{ij}^{\rm (th)}$ has a function form that is present in four-dimensions. Henceforth, $F_{ij}^{\rm (th)}$ is called the 4D-like phase space function. Meanwhile, no similarity to four-dimensional phase space is seen for $G_{ij}^{\rm (th)}$ and $H_{ij}^{\rm (th)}$, which are referred to as the 2D-specific phase space functions. Below, in the massless limit of down quark, we show that one SU(3) breaking combination that appears in Eq.~(\ref{Eq:7}) depends crucially on whether (a) $F_{ij}^{\rm (th)}$ is only included  or (b) all of $F_{ij}^{\rm (th)}$ $G_{ij}^{\rm (th)}$ and $H_{ij}^{\rm (th)}$ are considered in Eq.~(\ref{Eq:8}),
\begin{eqnarray}
&(\mathrm{a})& \left.\Gamma_{dd}^{(D^0,\: \mathrm{inc})}+\Gamma_{ss}^{(D^0,\: \mathrm{inc})}-2\Gamma_{sd}^{(D^0,\: \mathrm{inc})}\right|_{4\mathrm{D}-\mathrm{like}}=\Gamma^{(D^0,\: \mathrm{inc})}_{dd}[-2z_s^2+\mathcal{O}(z_s^3)],\label{Eq:10}\\
&(\mathrm{b})&\left.\Gamma_{dd}^{(D^0,\: \mathrm{inc})}+\Gamma_{ss}^{(D^0,\: \mathrm{inc})}-2\Gamma_{sd}^{(D^0,\: \mathrm{inc})}\right|_{4\mathrm{D}+2\mathrm{D}}=\Gamma^{(D^0,\: \mathrm{inc})}_{dd}[-2z_sR+\mathcal{O}(z_s^2)].\label{Eq:11}
\end{eqnarray}
Consequently, the case (a) is rather suppressed in the case of heavy charm quark. In view of this aspect, we shall show the final results for both (a) and (b) in Sec.~\ref{Sec:4}, by taking account of massive strange quark while treating down quark as a massless fermion.
\section{Numerical results}\label{Sec:4}
\begin{figure}[H]
\centering
\vspace{-2mm}
\includegraphics[width=9.7cm]{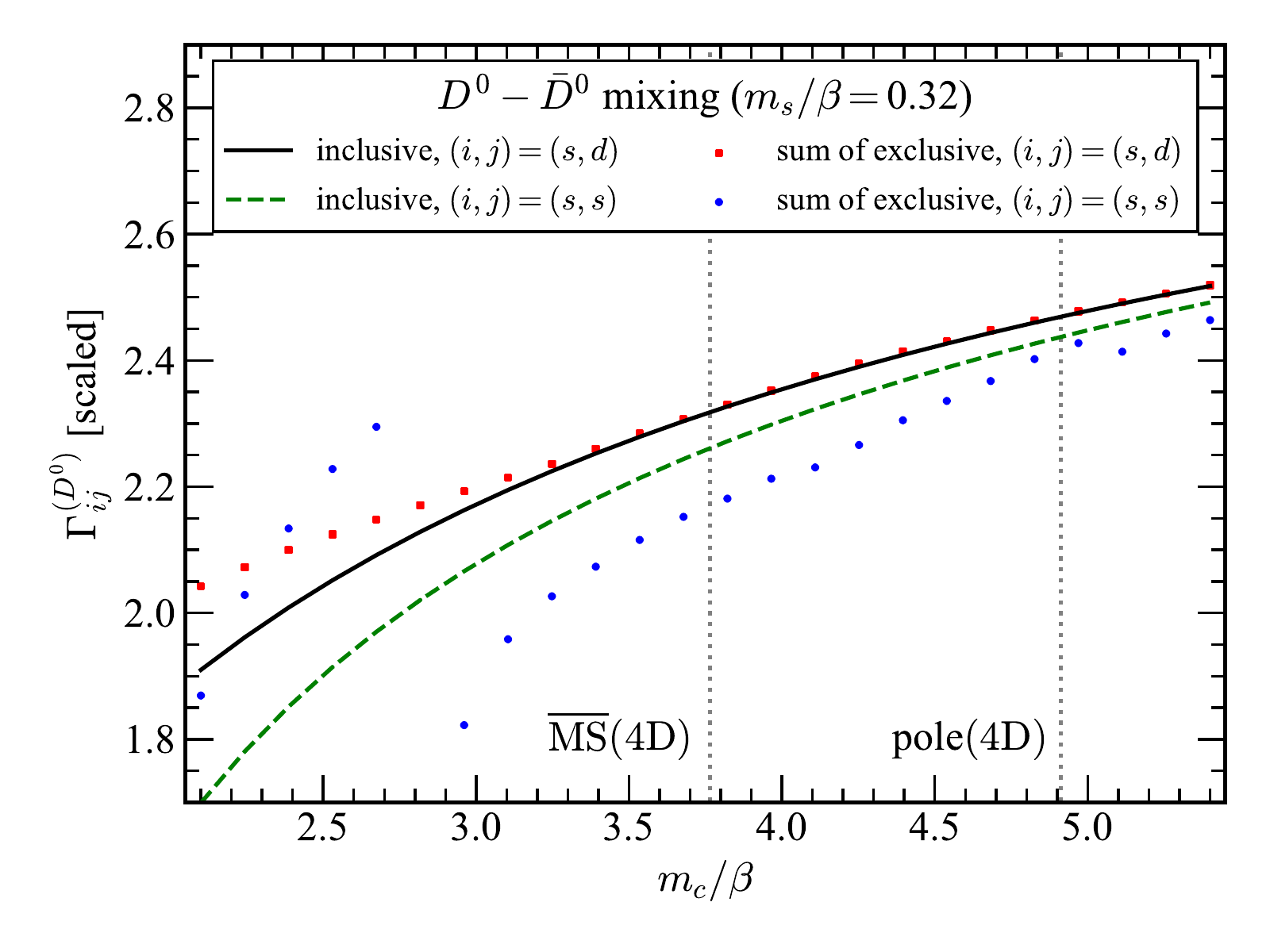}
\vspace{-6mm}
\caption{Exclusive and inclusive width differences in $D^0-\bar{D}^0$ mixing for individual flavor contributions. A black solid (green dashed) line stands for the inclusive result while red squares (blue points) show the sum of exclusive width differences for the intermediate $sd$ $(ss)$ contribution. Vertical grey dotted lines represent the $\overline{\mathrm{MS}}$ mass on the left and the pole mass on the right for charm quark in 4D. The vertical axis is given in the unit of $4N_cG_F^2(c_{\rm V}^2-c_{\rm A}^2)^2\beta/\pi$.}
\label{fig:1}
\end{figure}

In Fig.~\ref{fig:1}, the result in which exclusive $\Gamma_{ij}^{(D^0)}$  in Eq.~(\ref{Eq:12}) and inclusive one in Eq.~(\ref{Eq:8}) are compared for $(i, j)=(s, d), (s, s)$ is exhibited. The strange quark mass is set to the $\overline{\mathrm{MS}}$ mass at the scale of $m_c$ in 4D, corresponding to $m_s/\beta=0.32$. An obvious pattern of thresholds due to the two-dimensional phase space factor of the exclusive result in Eq.~(\ref{Eq:12}) is seen for $(i, j)=(s, s)$. As charm quark gets heavier, the exclusive and inclusive width differences asymptotically agree with one another for both $(i, j)=(s, d), (s, s)$. As for $(i, j)=(d, d)$, one can analytically confirm \cite{Umeeda:2021llf} that the two objects exactly agree in the massless limit of down quark.\par
\begin{figure}[H]
\centering
\vspace{-7.0mm}
\includegraphics[width=9.7cm]{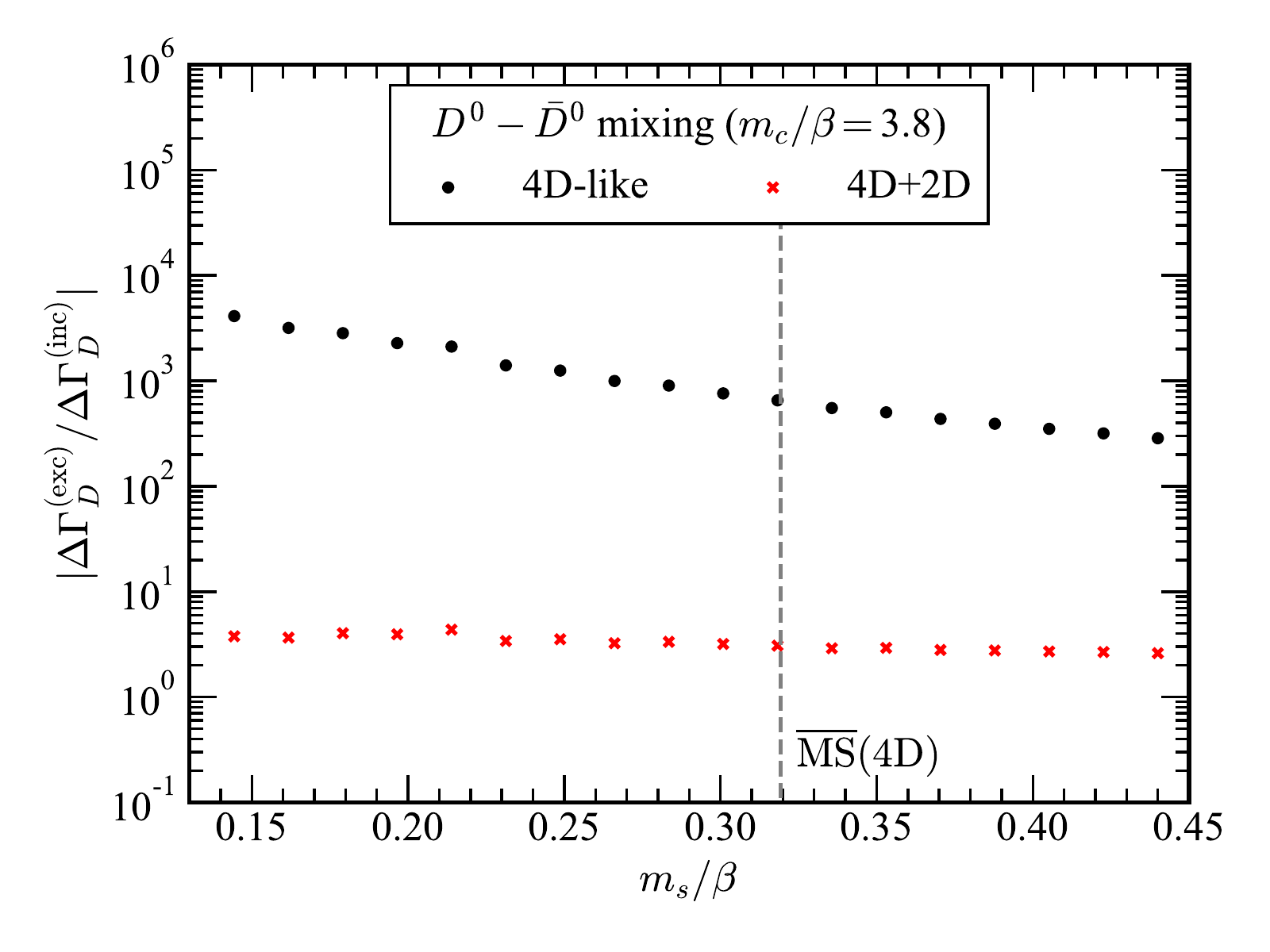}
\vspace{-5mm}
\caption{Absolute values for ratio of exclusive width difference to inclusive counterpart. Black points (red crosses) represent the case where the only 4D-like phase space function is included (all the possible phase space functions are contained). A vertical grey dashed line stands for the $\overline{\mathrm{MS}}$ mass for strange quark in 4D.}
\label{fig:2}
\end{figure}

In Fig.~\ref{fig:2}, the result for $|\Delta\Gamma_{D^0}^{\rm (exc)}/\Delta\Gamma_{D^0}^{\rm (inc)}|$ defined via Eq.~(\ref{Eq:6}) is shown. In this plot, we vary $m_s/\beta$ since this parameter significantly alters the order of magnitude as indicated for the inclusive analysis in Eqs.~(\ref{Eq:10}, \ref{Eq:11}). The charm quark mass is set to the $\overline{\mathrm{MS}}$ mass at the scale $m_c$ in 4D, corresponding to $m_c/\beta=3.8$. As was explained in Sec.~\ref{Sec:3.2}, the numerical results are shown for the two cases: (a) only 4D-like phase space is considered and (b) all the possible phase space functions are included in the inclusive analysis. As a result, we find that for $0.14<m_s/\beta<0.25$, the size of the exclusive width difference is larger than the inclusive counterpart by more than $10^3$ for (a).
To summarize, the exclusive and inclusive width differences for the $D^0-\bar{D}^0$ mixing are compared within the 't Hooft model in this work. The result has shown that the order of magnitude of the exclusive rate is larger than the OPE result from the four-quark operators, although the confirmed enhancement of $\mathcal{O}(10^3)$ is slightly smaller than what is indicated as the realistic observation in four-dimensions, that is, $y_{\rm ex}/y_{\rm th}\approx 10^{4}$. Therefore, it is shown within the model that the approximation where only the four-quark operators are included is not reasonable in the $D^0-\bar{D}^0$ mixing in sharp contrast to the $B^0_s-\bar{B}^0_s$ mixing, where the OPE result precisely agrees with the experimental value. Obviously, an evaluation for the contributions of six-quark and eight-quark operators, which entails a certain nonperturbative QCD method, is desirable in the four-dimensional analysis.
\acknowledgments
The author would like to thank the organizers of the CKM2021 for the active workshop in spite of the difficult pandemic times. Part of the analysis in this work was performed by the computational resources at Academia Sinica Grid Computing Centre (ASGC). This work was supported in part by MOST of R.O.C. under Grant No. MOST-110-2811-M-001-540-MY3.

\end{document}